\documentstyle[11pt,epsf]{article}
\def\be{\begin{equation}}
\def\bear{\begin{array}}
\def\eear{\end{array}}
\def\bea{\begin{eqnarray}}
\def\eea{\end{eqnarray}}

\def\ee{\end{equation}}
\def\c{\cite}

\begin{document}
\begin{center}
\LARGE {\bf Algebraic approach for shape invariant potentials in
Klein-Gordon equation}
\end{center}
\begin{center}
 {\bf $^{a}$M. R. Setare {\footnote {E-mail: rezakord@ipm.ir}}}
 {\bf $^{b}$O. Hatami {\footnote {E-mail:hatami{$\_$}physics@yahoo.com}}}\\
 {\it $^a$1 Department of Science, Payame Noor University, Bijar, Iran}\\
 {\it $^b$Department of Science, Payame Noor University, Sarpolzahab, Iran}\\
 \end{center}
\vskip 1.5cm
\begin{center}
{\bf{Abstract}}
 \end{center}
 The Shape invariant method has the algebraic structure and its algebras are
 infinite-dimensional. These algebras are converted into finite-dimensional under conditions.
  Based on the property of this method we obtain the algebraic structure of some physical potentials in the s-wave Klein-Gordon
  equation with scslar and vector potential which satisfy the Shape invariant condition.\\\vspace{1cm}\\
 {keywords: \it Klein-Gordon equation; Shape Invariance; Lie algebra.}
\\\vspace{8cm}\\
{$PACS$ $number:$ 03.65.Fd, 03.65.Ge, 02.20.Sv}\\
\newpage
\section{Introduction}In recent years there has been an
increasing interest in the study of relativistic wave equations,
particularly, the Klein-Gordon  equation \c{Greiner,Jana,Chen,de
Castro}, because the solutions of this equation play a fundamental
role in describing the relativistic effects in nuclear physics and
high energy physics, and other areas.
\\On the other hand, the supersymmetry quantum mechanics and shape
invariant method have an underly algebraic structure and the
associated Lie algebras are shown to be infinite dimensional
\c{Balantekin}. This method is an exact and elegant technique for
determining the eigenvalues and eigenfunctions of quantum mechanical
problems. Supersymmetry was first studied in the simplest case of
supersymmetric quantum mechanics by Witten \c{Witten}, Cooper and
Freeman \c{Cooper}. In 1983, the concept of shape invariant
potential was introduced by Gendenstein \c{Gendenshtein}, there has
been considerable discussion in the literature of this technique and
it has been applied to many of the solvable systems in quantum
mechanics \c{Cooper,Sukhatme}. It is well known that a potential is
said to be shape invariant, if its supersymmetry partner potential
has the same spatial dependence as the original potential with
possibly altered parameters, and the corresponding Hamiltonian can
be solved \c{Jana,Chen,Cooper,Sukhatme,Filho, aiza}.\\ In this
Letter we utilize the approach of Balantekin \c{Balantekin} for
shape invariant potentials. In section 2 of this paper we provide a
brief review of the shape invariance method and associated Lie
algebra. In section 3 we represent the Klein-Gordon equation with
equal scalar and vector five-parameter exponential-type potentials,
then we show a large class of exponential-type potentials can be
obtain by choosing the appropriate parameters in this potential. We
show that the Lie algebra of five-parameter exponential-type
potentials is $su(1,1)$, then in the same process we determine the
Lie algebra structure of Harmonic Oscillator with Heisenberg-Wely
algebra, and Morse potential with $su(1,1)$ algebra in the
Klein-Gordon equation. Finally, the conclusion is given in section
4.
\section{Supersymmetry quantum mechanics and algebraic formulation to shape invariance }
We note that supersymmetry method is based on the relation among the
energy spectrum, energy eigenfunction and phase shift of the
Hamiltonians associated with two supersymmetric partner potentials
$V_-(x)$ and $V_+(x)$ \c{Cooper}. Let us assume that the ground
state energy being zero, namely,
\begin{equation}
H_-(x)=H(x)-E_0,\label{1}\end{equation} where $E_0$ is the ground
state energy of $H(x)$, and introduce the following operators:
\begin{equation} \hat{A}={d\over{dx}}+W(x),\hspace{1cm}\hat{A^{\dag}}=-{d\over{dx}}+W(x),\label{2}\end{equation} where
$W(x)$ signifies the superpotential. From these two operators we can
construct two partner Hamiltonians
\begin{equation}\hat{H}_-=\hat{A^{\dag}}\hat{A}=-\frac{d^2}{dx^2}+V_-(x),\label{3}\end{equation}
\begin{equation}\hat{H}_+=\hat{A}\hat{A}^{\dag}=-\frac{d^2}{dx^2}+V_+(x).\label{4}\end{equation}
Since $\hat{A}$ annihilates the ground state
\begin{equation}\hat{A}\psi_0(x)=(\frac{d}{dx}+W(x))\psi_0(x)=0.\label{5}\end{equation}
The ground state wave function is obtained as:
\begin{equation}\psi_0(x)\sim\exp(-\int_{x_0}^x W(x)dx ),\label{6}\end{equation} where $x_0$
is an arbitrarily chosen reference point. A supersymmetric pair of
potentials\begin{equation}V_{\pm}(x;a_1)= W^2(x;a_1)\pm
W'(x;a_1),\label{7}\end{equation} are called shape invariant if
\begin{equation} V_{+}(x;a_1)=V_{-}(x;a_2)+ R(a_1),\label{8}\end{equation}
meaning that two supersymmetric partner potentials have the same
form, but are characterized by different values of parameters $a_1$
and $a_2$. The remainder $R(a_1)$ is independent of $x$ and the
parameter $a_2$ is a function of  the parameter $a_1$. So far two
classes of shape invariant are studied and discussed, in one class
the parameters are related to each other by a translation \c{B2,Ch}
\begin{equation}a_2=a_1+\eta\label{9}\end{equation}
and in second class they are related by a scaling \c{B3,A}
\begin{equation}a_2=qa_1.\label{10}\end{equation}In this paper we consider the
translation case. We define an operator which transforms the
parameters of partner potential \c{B2}
\begin{equation}\hat{T}(a_1)O(a_1)\hat{T}^{-1}(a_1)=O(a_2),\label{11}\end{equation}
where $\hat{T}(a_1)$ and  $\hat{T}^{-1}(a_1)$ is as follows
\begin{equation}\hat{T}(a_1)=\mathrm{exp}(\eta\frac{\partial}{\partial
a_1}),\hspace{1.5cm}\hat{T}^{-1}(a_1)=\hat{T^\dag}(a_1)=\mathrm{exp}(-\eta\frac{\partial}{\partial
a_1}).\label{12}\end{equation} We introduce new operators in the
following form to establish the algebraic structure
\begin{equation}\hat{B}_+=\hat{A^{\dag}}(a_1)\hat{T}(a_1),\label{13}\end{equation}
\begin{equation}\hspace{1cm}\hat{B}_-=\hat{B}^\dag_+=\hat{T}^\dag(a_1)\hat{A}(a_1).\label{14}\end{equation}
So the Hamiltonians of Eqs. (3), (4) can be written as
\begin{equation}\hat{H}_-=\hat{A^{\dag}}\hat{A}=\hat{B}_+\hat{B}_{-},
\hspace
{2cm}\hat{H}_+=\hat{A}\hat{A^{\dag}}=\hat{T}\hat{B}_-\hat{B}_{+}\hat{T}^\dag.\label{15}\end{equation}
Eq. (8) can be written as a commutator
\begin{equation}[\hat{B}_-,\hat{B}_{+}]=\hat{T}^\dag(a_1)R(a_1)\hat{T}(a_1)\equiv R(a_0),\label{16}\end{equation}
where we defined \begin{equation}
a_n=a_1+(n-1)\eta\label{17}\end{equation} and we have used
\begin{equation}R(a_n) =\hat{T}(a_1 )R(a_{n-1})\hat{T}^\dag(a_1),\label{18}\end{equation}
valid for any $n$. We defined the commutation relation (16) for
determine the Lie algebra associated by the shape-invariant and also
this equation shows that $\hat{B}_-$ and $\hat{B}_{+}$ are the
appropriate creation and annihilation operators for the spectra of
the shape-invariant potentials provided that their non-commutativity
with $R(a_0)$ is taken into account. The additional relations
\begin{equation}R(a_n)\hat{B}_{+}=\hat{B}_{+}R(a_{n-1}),\label{19}\end{equation}
\begin{equation}R(a_n)\hat{B}_{-}=\hat{B}_{-}R(a_{n+1})\label{20}\end{equation}
which readily follow from Eqs. (13), (14), (18). One can show the
following relations
\begin{equation}\hspace{-3.75cm}\hat{B}_{-}|\psi_0\rangle=\hat{A}|\psi_0\rangle=0,\label{21}\end{equation}
\begin{equation}\hspace{-.25cm}[\hat{H},\hat{B}^n_{+}]=(R(a_1)+R(a_2)+··+R(a_n))\hat{B}^n_{+},\label{22}\end{equation}
\begin{equation}\hspace{0.3cm}[\hat{H},\hat{B}^n_{-}]=-\hat{B}^n_{-}(R(a_1)+R(a_2)+..+R(a_n)),\label{23}\end{equation}
meaning that $\hat{B}^{n}_+|\psi_0\rangle$ is an eigenstate of the
Hamiltonian with the eigenvalue $R(a_1)+ R(a_2)+..+R(a_n)$. The
normalized wavefunction is
\begin{equation}|\psi_n\rangle=\frac{1}{\sqrt{R(a_1)+ R(a_2)+··+R(a_n)}}\hat{B}_{+}..
\frac{1}{\sqrt{R(a_1)+R(a_2)}}\hat{B}_{+}\frac{1}{\sqrt{R(a_1)}}\hat{B}_{+}|\psi_0\rangle,\label{24}\end{equation}
where $|\psi_0\rangle$ depends on $x$ and parameter $a_1$. It is
easily to show the commutation relations
\begin{equation}[\hat{B}_{+},R(a_0)]=(R(a_1)- R(a_0))\hat{B}_{+},\label{25}\end{equation}
\begin{equation}[\hat{B}_{+},(R(a_1)- R(a_0))\hat{B}_{+}]=(R(a_2)-2R(a_1)+R(a_0))\hat{B}^2_{+},\label{26}\end{equation}
and so on. There is an infinite number of such commutation
relations. If $R(a_n)$ satisfy certain relations one of the
commutators in this series may vanish. These commutation relations
with their complex conjugates form a Lie algebra with a finite
number of elements. We know that for confining potentials the $n$'th
eigenvalue for large $n$ obeys the constraint \c{m}
\begin{equation}E_n\leq constant\times n^2.\label{27}\end{equation}
We consider some potentials where have a finite algebra, for these
potentials $E_n$ is as
\begin{equation}E_n=\mu n^2+\nu n+\kappa,\label{28}\end{equation}
where $\mu,$ $\nu,$ $\kappa$ are constants. By using Eq. (28) one
can obtain
\begin{equation}R(a_n)=E_n-E_{n-1}=2\mu n+\nu-\mu\label{29}\end{equation}
then we have
\begin{equation}R(a_n)-R(a_{n-1})=2\mu.\label{30}\end{equation}
So, the Lie algebra is finite
\begin{equation}[\hat{B}_{-},\hat{B}_{+}]=R(a_0),\hspace {1cm}[\hat{B}_{-},R(a_0)]=-2\mu\hat{B}_{-},\hspace {1cm}[\hat{B}_{+},R(a_0)]=2\mu\hat{B}_{+}.\label{31}\end{equation}
For the case $\mu=0$ it is Heisenberg-Wely algebra, if $\mu\neq0$
the algebra is either $SU(1,1)$ or $SU(2)$ depends on the sign of
$\mu$. The energy spectrum corresponding to $H(x)$ can be derived
from the shape-invariance conditions
\begin{equation} E_{n}=\sum^{n}_{k=1}R(a_k)+E_{0} \hspace{1cm}.\label{32}\end{equation}
\section{EXAPLES}Now we would like to apply the discussion in the
previous section for some potentials which satisfy the shape
invariant condition, then we determine the Lie algebra of these
potentials.
\subsection{Five-parameter exponential-type potential model}
The s-wave Klein-Gordon equation with scalar and vector potentials
is ($\hbar=c=1$) \c{Greiner}
\begin{equation}\{\frac{d^2}{dr^2}+[E-V(r)]^2-[M+S(r)]^2\}u(r)=0,\label{33}\end{equation}
where the radial wave function is $R(r)=u(r)/{r}$ and $M$ is the
mass, $E$ is the energy and $S(r)$ and $V(r)$ denote the scalar and
vector potentials, respectively. If $S(r)\geq V(r)$, there exists
bound state solution. In the case of equal scalar and vector
potentials, $S(r)=V(r)$, Eq. (1)
reduces to a Schr\"{o}dinger-like equation\\
\begin{equation}\{-\frac{d^2}{dr^2}+2(M+E)V(r)\}u(r)=(E^2-M^2)u(r).\label{34}\end{equation}
We consider the five-parameter exponential-type potential model,
which is written in the form of \c{c}
\begin{equation}V(r)=\frac{Q^{2}_3+\mathrm{g}-\frac{Q_{2}^2}{q}+2\alpha
Q_2}{e^{2\alpha r}+q} +\frac{Q_2^2-qQ^2_3-2\alpha qQ_2}{(e^{2\alpha
r}+q)^2}+\frac{\mathrm{g}\frac{Q_3}{Q_2}-\frac{Q_2Q_3}{q}+\alpha
Q_3}{e^{2\alpha r}+q}e^{\alpha r}+\frac{2Q_2Q_3-2\alpha
qQ_3}{(e^{2\alpha r}+q)^2}e^{\alpha r},\label{35}\end{equation}
where the range of parameter $q$ is $-1\leq q<0$ or $q >0$.
Combining Eqs. (34), (35) leads to
\begin{equation}\begin{array}{c}\hspace{-7cm}\{-\frac{d^2}{dr^2}+2(m+E)
\{\frac{Q^{2}_3+\mathrm{g}-\frac{Q_{2}^2}{q}+2\alpha Q_2}{e^{2\alpha
r}+q} +\frac{Q_2^2-qQ^2_3-2\alpha qQ_2}{(e^{2\alpha
r}+q)^2}\\\\\hspace{2cm}+\frac{\mathrm{g}\frac{Q_3}{Q_2}-\frac{Q_2Q_3}{q}+\alpha
Q_3}{e^{2\alpha r}+q}e^{\alpha r}+\frac{2Q_2Q_3-2\alpha
qQ_3}{(e^{2\alpha r}+q)^2}e^{\alpha r}\}u(r)=(E^2-m^2)u(r).
\end{array} \label{36}\end{equation}\\ The corresponding superpotential is
\begin{equation}W(r,a_n)=\frac{\mathrm{g}}{2a_n}-\frac{a_n}{2q}
+\frac{a_n}{e^{2\alpha r}+q}+\frac{Q_3}{e^{2\alpha r}+q}e^{\alpha
r},\label{37}\end{equation}where
\begin{equation}a_n=Q_2+2n\alpha q.\label{38}\end{equation}The remainder in Eq. (8) is given by
\begin{equation}R(a_n)=-[\frac{\mathrm{g}}{2(Q_2+2n\alpha q)}-\frac{(Q_2+2n\alpha q)}{2q}]^2
+[\frac{\mathrm{g}}{2(Q_2+2(n-1)\alpha q)}-\frac{(Q_2+2(n-1)\alpha
q)}{2q}]^2. \label{39}\end{equation}For the case $\mathrm{g}=0$ we
have
\begin{equation}R(a_n)=[\frac{(Q_2+2(n-1)\alpha q)}{2q}]^2-[\frac{(Q_2+2n\alpha q)}{2q}]^2=\alpha^2(1-2n)-\alpha
\frac{Q_2}{q},\label{40}\end{equation}which gives
\begin{equation}R(a_n)-R(a_{n-1})=-2\alpha^2,\label{41}\end{equation}
then by using Eq. (40) we have
\begin{equation}R(a_0)=\alpha(\alpha-\frac{Q_2}{q}).\label{42}\end{equation}
We define the following operators
\begin{equation}K_0=\frac{1}{2\alpha^2}R(a_0),\hspace{2cm}K_\pm=\frac{1}{\alpha}B_\pm,\label{43}\end{equation}
Using Eqs. (41), (43) one can show that the shape invariant algebra
for the five-parameter exponential-type potential model is $su(1,1)$
\begin{equation}[K_+,K_-]=-2K_0,\hspace{1cm}[K_0,K_-]=-K_-,\hspace{1cm}[K_0,K_+]=K_+\label{44}\end{equation}
\\\textbf{{Discussion}}\\
By choosing appropriate parameters in the five-parameter model
exponential-type potential one can construct seven typical potential \c{y, X, jia, part}\\\\
\textit{\textbf{Case:1.}} $V_0\tanh^2(\frac{r}{d})$
\textit{\textbf{potential}}\\\\
Choosing $q=1$, $\alpha=\frac{1}{d}$, $\mathrm{g}=0$, $Q_3=0$, and
set
\begin{equation}Q_2^2-2\alpha Q_2=4V_0,\label{45}\end{equation} then the five-parameter exponential-type
potential becomes the $V_0\tanh^2(\frac{r}{d})$ potential with
constant shift.
\begin{equation}V(r)=V_0\tanh^2(\frac{r}{d})-V_0.\label{46}\end{equation}
\\\\
\textit{\textbf{Case:2. Scarf II potential}}\\ \\If $q=1$,
$\mathrm{g}=0$, and put
\begin{equation}Q_2^2-Q_3^2-2\alpha Q_2=-4[B^2-A(A+\alpha)],\label{47}\end{equation}
\begin{equation}Q_2Q_3+\alpha Q_3=-2B(2A+\alpha),\label{48}\end{equation}\\ the potential (35) turns to the
Scarf II potential
\begin{equation}V(r)=[B^2-A(A+\alpha)]{\textrm{sech}}^2{\alpha r}+B(2A+\alpha){\textrm{sech}}{\alpha r}\tanh{\alpha
r},\label{49}\end{equation}\\
\textit{\textbf{Case:3. Generalized P\"{o}schl-Teller}}\\\\Choosing
$q=-1$, $\mathrm{g}=0$, and setting
\begin{equation}Q_2^2+Q_3^2+2\alpha Q_2=4[B^2+A(A+\alpha)],\label{50}\end{equation}
\begin{equation}Q_2Q_3+\alpha Q_3=-2B(2A+\alpha),\label{51}\end{equation}
the five-parameter exponential-type potential becomes the
generalized p\"{o}schl-Teller potential
\begin{equation}V(r)=[B^2+A(A+\alpha)]{\textrm{cosech}}^2{\alpha r}-B(2A+\alpha)\textrm{cosech}{\alpha
r}\coth{\alpha r},\label{52}\end{equation}\\
\textit{\textbf{Case:4. P\"{o}schl-Teller II potential}}\\\\If we
choose $q=-1$, $\mathrm{g}=0$, $\alpha\rightarrow 2\alpha$ and put
\begin{equation}Q_2^2+Q_3^2+4\alpha Q_2=8[A(A+\alpha)+B(B-\alpha)],\label{53}\end{equation}
\begin{equation}Q_2Q_3+2\alpha Q_3=-4[A(A+\alpha)-B(B-\alpha)],\label{54}\end{equation}
the five-parameter exponential-type potential turns to the
p\"{o}schl-Teller II potential potential
\begin{equation}V(r)=-A(A+\alpha){\textrm{sech}}^2{\alpha r}+B(B-\alpha)\textrm{cosech}^2{\alpha r},\label{55}\end{equation}\\
\textit{\textbf{Case:5. PT-symmetric version of the Scarf II.
potential}}\\\\Choosing $q=1$, $\mathrm{g}=0$, and letting
\begin{equation}Q_2^2-Q_3^2-2\alpha Q_2=4V_1,\label{56}\end{equation}
\begin{equation}Q_2Q_3-\alpha Q_3=2iV_2,\label{57}\end{equation}
the five-parameter exponential-type potential yields the
PT-symmetric version of the Scarf II potential
\begin{equation}V(r)=-V_1{\textrm{sech}}^2{\alpha r}-iV_2 \textrm{sech}{\alpha r}\tanh{\alpha r},\label{58}\end{equation}\\
\textit{\textbf{Case:6. Symmetrical double-well potential}} \\\\If
we take $q=1$, $\mathrm{g}=0$, $Q_3=0$ and set
\begin{equation}Q_2^2-2\alpha Q_2=4(V_1+V_2),\label{59}\end{equation}
Eq. (35) reduce to the standard symmetrical double-well potential
with constant shift.\c{f}
\begin{equation}V(r)=V_1\tanh^2{\alpha r}-\frac{V_2}{\cosh^2{\alpha r}}-V_1.\label{60}\end{equation}
\textit{\textbf{Case:7. Reflectionless-type potential}}\\ \\If we
define $q=1$, $\alpha=1$, $\mathrm{g}=0$, $Q_3=0$ and put
\begin{equation}Q_2^2-2\alpha Q_2=2\lambda(\lambda+1),\label{61}\end{equation}
Eq. (35) turns to the reflectionless-type potential
\begin{equation}V(r)=-\frac{1}{2}\lambda(\lambda+1)\textrm{sech}^2{r}.\label{62}\end{equation}
\subsection{Harmonic Oscillator}
We take the scalar and vector potentials  as follows:
\begin{equation}V(r)=\frac{V_0}{2}r^2.\label{63}\end{equation}
Inserting these potentials into Eq. (34) we obtain
\begin{equation}\{-\frac{d^2}{dr^2}+(M+E)V_0r^2\}u(r)=(E^2-M^2)u(r).\label{64}\end{equation}
The superpotential is
\begin{equation}W(r,a_n)=a_n r,\label{65}\end{equation}
where
\begin{equation}a_n^2=a_1^2=(M+E)V_0.\label{66}\end{equation}
The remainder in Eq. (8) is given by
\begin{equation}R(a_n)=2a_n=2\sqrt{(M+E)V_0}\label{67}\end{equation}
Introducing the following operators
\begin{equation}K_0=\frac{R(a_0)}{\sqrt{(M+E)V_0}}=I,\hspace{2cm}K_\pm=\frac{1}{{\sqrt{(M+E)V_0}}}B_\pm,\label{68}\end{equation}
so, we have
\begin{equation}[K_-,K_+]=I,\hspace{1cm}[K_0,K_\pm]=0,\hspace{1cm}[K_\pm,K_+K_-]=\mp K_\pm,\label{69}\end{equation}
where these commutation relations denote the Heisenberg-Wely
algebra.
\subsection{Morse potential}
We consider scalar and vector  exponential-type potentials written,
respectively, as follows:
\begin{equation}S(r)=S_0e^{-\alpha r},\hspace{2cm} V(r)=V_0e^{-\alpha r},\label{70}\end{equation}
substituting these potentials into Eq. (33) we have
\begin{equation}\{-\frac{d^2}{dr^2}+(S^2_0-V_0^2)e^{-2\alpha r}+2(MS_0+EV_0)e^{-\alpha r}\}u(r)=(E^2-M^2)u(r).\label{71}\end{equation}
which is called Klein-Gordon-Morse potential \c{Chen}. The
superpotential is
\begin{equation}W(r,a_n)=a_n-\sqrt{S^2_0-V_0^2}\hspace{0.2cm}e^{-\alpha r},\label{72}\end{equation}
where
\begin{equation}a_n=a_1-(n-1)\alpha,\hspace{2cm}a_1=-\frac{\alpha}{2}+\frac{MS_0+EV_0}{\sqrt{S^2_0-V_0^2}}.\label{73}\end{equation}
The remainder is given by
\begin{equation}R(a_n)=(2a_n-\alpha)\alpha,\label{74}\end{equation}
We introduce the following operators
\begin{equation}K_0=\frac{R(a_0)}{2\alpha^2},\hspace{2cm}K_\pm=\frac{B_\pm}{\alpha}\label{75}\end{equation}
where
\begin{equation}R(a_0)=\frac{2(MS_0+EV_0)}{\sqrt{S^2_0-V_0^2}}.\label{76}\end{equation}
From Eq. (75) one can obtain that the shape invariant algebra for
morse potential in Klein-Gordon equation to be $su(1,1)$ algebra
\begin{equation}[K_+,K_-]=-2K_0,\hspace{2cm}[K_0,K_\pm]=\pm K_\pm\label{77}\end{equation}

\section{Conclusion}In this paper, we have briefly recalled the concepts of shape invariance method
and its algebra structure. The shape invariant algebras, in general,
have infinite-dimensional and under conditions they become
finite-dimensional.  We have considered the three-dimensional s-wave
relativistic Klein-Gordon equation with equal scalar and vector
five-parameter exponential-type potential. This equation satisfied
the Shape invariant condition, then we showed a large class of
exponential-type potentials can be obtain by choosing the
appropriate parameters in this potential. By using the shape
invariant algebra we have shown that the Lie algebra of
five-parameter exponential-type potentials is $su(1,1)$. Then we
have considered the Harmonic oscillator potential and we found that
this potential has the Heisenberg-Wely algebra. In the same process,
we determined the Lie algebra structure of Morse potential with
$su(1,1)$ algebra in the Klein-Gordon equation.

\end{document}